\documentclass[useAMS,usenatbib]{mn2e}
\usepackage{epsf}
\title{Lyman Break Galaxies at $z \sim 1$ and the evolution of dust attenuation in star-forming galaxies the redshift}

\author[D. Burgarella, E. Le Floc'h, T. T. Takeuchi, J.S. Huang, V. Buat, G.H. Rieke and K.D. Tyler]{D. Burgarella\thanks{E-mail:denis.burgarella@oamp.fr}, E. Le Floc'h\thanks{E-mail:elefloch@as.arizona.edu}, T.T. Takeuchi\thanks{E-mail:takeuchi@iar.nagoya-u.ac.jp}, J.S. Huang\thanks{E-mail:jhuang@cfa.harvard.edu}, V. Buat\thanks{E-mail:veronique.buat@oamp.fr}, G.H. Rieke\thanks{E-mail:grieke@as.arizona.edu} \and K.D. Tyler\thanks{E-mail:ktyler@as.arizona.edu}}

\begin{document}

\date{Accepted. Received; in original form }

\pagerange{\pageref{firstpage}--\pageref{lastpage}} \pubyear{2006}

\maketitle

\label{firstpage}

\begin{abstract}
Ultraviolet (UV) galaxies have been selected from the $GALEX$ deep imaging survey. The presence of a FUV-dropout 
in their spectral energy distributions proved to be a very complete (83.3 \%) but not very efficient (21.4 \%)  
tool for identifying Lyman Break Galaxies (LBGs) at $z\sim 1$. In this paper, we explore the physical 
properties of these galaxies and how they contribute to the total star formation rate. We divide the LBG sample 
into two sub-classes:  red LBGs (RLBGs) detected at $\lambda = 24 \mu m$ which are mainly Luminous IR Galaxies 
(LIRGs) and blue LBGs (BLBGs) undetected at $\lambda = 24 \mu m$ down to the MIPS/GTO limiting flux density of 
83 $\mu Jy$. Two of the RLBGs are also detected at 70 $\mu m$. The median SED of the RLBGs is similar (above $\lambda 
\sim 1 \mu m$) to that of a luminous dusty starburst at $z \sim 1.44$, HR10. However, unlike local Luminous 
and Ultra Luminous IR Galaxies, RLBGs are UV bright objects. We suggest that these objects contain a large 
amount of dust but that some bare stellar populations are also directly visible. The median SED of the BLBGs is 
consistent with their containing the same stellar population as the RLBGs (i.e. a 250 - 500 Myrs old, exponentially 
decaying star formation history) but with a lower dust content.
The luminosity function of our LBG sample at $z \sim 1$ is similar to the luminosity function of NUV-selected 
galaxies at the same redshift. The integrated luminosity densities of $z \sim 1$ LBGs and NUV-selected galaxies 
are very consistent. 
Making use of the RLBG sample, we show that star formation rates (SFRs) estimated from UV measurements and 
corrected using the $IRX-\beta$ method provide average total $SFR_{TOT}$ in agreement with the sum of the UV 
and infrared contributions: $SFR_{UV} + SFR_{dust}$. However, $IRX-\beta$-based $SFR_{TOT}$ shows a large dispersion.
Summing up the detected UV ($1150 \AA$ rest-frame) and IR-based star formation rates of the detected objects, 
we find that only one third of the total (i.e. UV + dust) LBG SFR resides in BLBGs and two thirds 
in RLBGs, even though most LBGs at $z \sim 1$ are BLBGs. On the other hand, the total SFR of LBGs accounts 
for only 11\% of the total SFR at $z \sim 1$.
Finally, we observe a regular decrease of the luminosity ratio $L_{dust} / L_{FUV}$ from $z = 0$ to 
$z \approx 2$ for UV-selected samples.
\end{abstract}


\section{Introduction}

The evolution of the star formation density with redshift contains fundamental information from the 
perspectives both of galaxy evolution and of cosmology. At intermediate redshift, star formation rates (SFRs)
are estimated through a combination of UV, mid-IR, radio, and emission line techniques. 
Lyman Break Galaxies (LBGs) are used to estimate SFRs at high redshift. They are assumed to 
be related to FUV-selected samples selected via a color-color plane (e.g. Bouwens et al. 2006, 
Reddy et al. 2006). A major issue in determining the star formation density is to cross calibrate the 
several tracers used to estimate the SFRs. This paper tests whether the star formation rate 
estimated from the far ultraviolet (FUV) continuum (UV SFR or SFR$_{UV}$) of UV-selected galaxies 
is equivalent to that from the infrared luminosity from dust estimated by integrating over the 
wavelength range $1 - 1000 \mu m$ (dust SFR or SFR$_{dust}$, equivalently SFR$_{TIR}$).

Even though we know that some dust is present in some LBGs (e.g. 
Chapman et al. 2002), it is difficult to get a direct estimate of how much in a statistically representative way 
because only a few LBGs at $z > 2$ have been detected in the FIR and in the sub millimeter (submm) 
ranges (e.g. Chapman et al. 2000). Even $Spitzer$ has not detected a large sample of $z > 2$ LBGs 
in the regime where the dust emission is the main component of the spectrum (e.g. Huang et al. 2005).
Adelberger \& Steidel (2000) concluded that the bulk of the submm background is 
produced by moderately obscured galaxies similar to the ones already detected in UV-selected 
surveys. Most of the dust-to-FUV luminosity ratios ($L_{dust} / L_{FUV}$) for these galaxies are estimated from 
the observed optical and by using the UV slope vs. $L_{dust} / L_{FUV}$ (IRX-$\beta$) relation 
defined by Meurer, Heckman \& Calzetti (1999). However, the validity of this law is debated 
(Bell 2002, Buat et al. 2005, Goldader et al. 2002). Nevertheless, using this law, dust luminosities 
are often estimated without FIR data (e.g. Adelberger \& Steidel. 2000). Can we trust these results? 
Sawicki (2001) concluded that "it is unlikely that a consensus about the amount of stellar light 
intercepted by dust will be reached on the basis of rest-frame UV and optical data alone".

We therefore need rest-frame FIR data to advance our knowledge of the dust content of UV-selected 
galaxies. In this paper we approach this goal by observing LBGs at $z \sim 1$ where we can directly evaluate the 
dust luminosity from $24 \mu m$ data. Burgarella et al. (2006b) studied an initial sample of LBGs in the 
redshift range $0.9 \leq z \leq 1.3$ identified from $GALEX$ data in the 
ultraviolet; a portion of this sample was also detected with 
$Spitzer/MIPS$ in the IR. The remainder are below the limiting MIPS flux density of $83 \mu Jy$. 
Although this is a true Lyman Break sample at $z \sim 1$, the selection is different from LBGs at $z \ge 3$ 
because the effect of the intergalactic medium is much lower at $z \sim 1$ than at higher redshift. 
The same phenomenon applies, although at a lower level, for BM/BX galaxies selected at $z \sim 2$ 
(e.g. Reddy et al. 2006). Thus, although exact analogs of the high redshift LBGs cannot be
used, the main objective in the previous work was to define a rest-frame FUV-selected sample with a 
measured estimate of the dust luminosity from IR data.

In this paper, we expand upon this approach. We describe the selection of 
the sample in Section 2. Section 3 is devoted to the building and the analysis of 
the spectral energy distributions (SEDs) of LBGs (i.e. red LBGs (RLBGs) and blue LBGs (BLBGs)) 
and to their comparison to the SEDs of LBGs at $z \approx 3$. Then, in Section 4, we will derive 
the luminosity function of the $z \sim 1$ LBG sample and compare it to the luminosity function 
of an ultraviolet-selected sample. In Section 5, we will address the question of the total star 
formation density at $z \sim 1$ and whether this total star formation density can be evaluated 
from an ultraviolet-selected sample, an infrared-selected sample or whether both of them are required. 
In Section 6, we will study the evolution of the $L_{dust} / L_{FUV}$ ratio for UV-selected samples or, in other 
words, of the evolution of the dust attenuation with redshift. We will finish this paper in 
Section 7 by estimating how we can detect LBGs in the infrared with facilities planned for the next decade.

We assume a cosmology with $H_0 = 70 ~km.s^{-1}.Mpc^{-1}$, $\Omega_M = 0.3$ and $\Omega_{VAC}=0.7$ in this paper.

\section{Definition of the UV and LBG Samples}

Our first paper (Burgarella et al. 2006b) on the LBG sample at $z \sim 1$ used GALEX pipeline data. 
Here, we use the deeper (76444 sec) NUV image available in GALEX Release 2 (GR2) whose central 
coordinates are $\alpha$ = 03h32m30.7s, and $\delta$ = -27deg52'16.9'' (J2000.0). An analysis of 
GALEX deep images shows that galaxies overlap, producing an apparent brightening of some of the detected 
objects in the tables available from the Multi Mission Archive at STScI (MAST). This effect starts to 
apply to objects brighter than $NUV \approx 22.0$. 

To deal with these effects systematically, we put all of our images at a common resolution
and extracted photometry by means of point spread function fitting. The spatial resolution of the GALEX 
images was slightly degraded by applying a median 3x3 filtering to the FUV and NUV images. 
The GALEX "beam" is taken to be the solid angle of the 1$\sigma$ radius circle of a 
Gaussian point-spread function (PSF) : $\Omega_{beam} = \pi \sigma^2$ as defined in Hogg (2001). Note that 
for our GALEX analysis, $\theta_{FWHM} = 6$ arcsec which translates into $\sigma = \theta_{FWHM} / 2.35 = 
2.55$ arcsec. The final angular resolution in both bands therefore, matches the 
MIPS $24 \mu m$ angular resolution. 

Confusion noise is significant in these images (see Dole et al. 2004 for 24$\mu$m). 
The size of the filtered GALEX beam $\Omega_{beam} = 4.81 \times 10^{-10}$ sr. 
Since the total field of view of GALEX corresponds to $\Omega_{GALEX} \sim 3.45 \times 10^{-4}$ sr, 
we obtain a confusion limit at $3 \sigma$:

$$ (s/b)_{conf}^{3\sigma} = {{N_{sources}} \over {(~\Omega_{GALEX} / \Omega_{beam})}}= 0.063 {\rm ~source~per~beam}$$

\noindent
or $\sim$ 16 beams per source at the 80\% completeness limit defined below for the NUV-selected sample.  This is deeper than the criterion published by Hogg (2001) of 20 beams per source but shallower than the $\sim$ 12 beams per source estimated by Dole et al. (2004) and Jeong et al. (2006). Takeuchi \& Ishii (2004) have published a relation to estimate the confusion limit at $5 \sigma$. Our number counts can be fitted by a power law with a slope $\gamma 
\approx 1.45$. Using $\varepsilon=2$ and $a=5$ as recommended by Takeuchi \& Ishii (2004), we can compute:

$$(s/b)_{conf}^{5\sigma} = {{\varepsilon^2 (2 - \gamma)} \over {2 a^2}} = 0.044 {\rm ~source~per~beam}$$

\noindent
or 23 beams per source. Accounting for the difference in detection level ($3 \sigma$ for the former 
and $5 \sigma$ for the latter), the two values are consistent.

We used DAOPHOT for photometry in these images with a PSF built 
from 10 bright point sources in the FUV and NUV. DAOPHOT was developed for point sources. Le Floc'h et al. 
(2005) show it is suitable to measure $24 \mu m$ flux densities of Spitzer/MIPS sources. 
Our UV sources are restricted to the redshift range $0.9 \le z \le 1.3$. De Mello et 
al. (2006) showed that UV-selected galaxies in the redshift range $0.8 \le z \le 1.3$ have $R_e = 1.6 \pm 0.4$ 
kpc i.e. FWHM $\approx 6$ kpc. At $z \sim 1$ this size corresponds to 0.7 arcsec. This is well below the 
angular resolution of our images, so we can safely use DAOPHOT for flux estimation in the targeted redshift range.  

We detect about $10^5$ objects in the NUV down to 3 $\sigma$ with the DAOPHOT DAOFIND task, which is in agreement with number counts by Xu et al. (2005). Using the DAOPHOT ADDSTAR task, we estimate the 80\% completeness limit to be NUV=26.2 for the NUV-selected sample. The NUV-selected sample used hereafter 
corresponds to all objects with a NUV magnitude $NUV \le 26.2$; it contains 45366 objects in the GALEX field of view but only part of them have a COMBO 17 counterpart (i.e. a redshift) mainly because most GALEX objects lie outside the COMBO 17 field of view. This source density corresponds to $\sim$ 16 beams per source, that is, just at the 3-$\sigma$ confusion limit. 

To check the validity of our approach, we also apply DAOPHOT PHOT aperture photometry (integrated 
over 3 arcsec with an aperture correction estimated from the PSF) to a set of isolated objects (i.e. no 
COMBO 17 counterparts within a 4-arcsec radius) that we compare to DAOPHOT magnitudes for the same objects. 
The difference in magnitude (DAOPHOT - PHOT) amounts to $0.17 \pm 0.26$ which is about the uncertainty of 
a typical GALEX measurement (Morrissey et al. 2005) given that we reach very faint magnitudes, confirming that 
DAOPHOT provides good estimates for the GALEX magnitudes.

\begin{table*}
\centering
%\begin{maxipage}{80mm}
\caption{Ultraviolet-selected sample down to NUV=26.2 within COMBO 17 (C17) field of view with magnitudes 
estimated from DAOPHOT. We stress that the magnitudes of resolved objects must be used with care because 
DAOPHOT is designed for point sources. However, for galaxies beyond $z \approx 0.5$, DAOPHOT photometry is 
valid (see text for a more detailed discussion).  46321 rows for 45366 $id_{UV}$ are listed out of which 38057 correspond to GALEX objects without COMBO 17 counterparts, 6404 GALEX objects have a single C17 counterpart and a few ones with up to 4 counterparts (859 $\times$ 2, 42 $\times$ 2 and 4 $\times$ 4). Several rows with the same id$_{UV}$ are listed whenever several C17 counterparts are found. For each row, columns give (1) the number of C17 counterparts (1), the UV id (2) and coordinate (3,4), the DAOPHOT FUV and NUV ABmag from PSF fitting with their estimated uncertainties (5-8); the C17 id (9), the R magnitude and uncertainty (10,11), the photometric redshift from C17 with a quality flag (0 means from $MC\_z\_ml$ and (1) from $MC\_z$ in C17) and finally the magnitude estimated from PHOT (3 arcsec aperture).}

\begin{tabular}{@{}ccccccccccccccc@{}}
\hline
n$_C17$ & id$_{UV}$   & RA$_{UV}$ & Dec$_{UV}$ & FUV & err$_{FUV}$ & NUV & err$_{NUV}$ & id$_{C17}$   &  R$_{mag}$ & err$_{R_{mag}}$ & z & z$_{qal}$ & NUV$_{phot3}$ & err$_{NUV_{phot3}}$ \\
\hline
\hline
1	& 41129	& 52.946268	& -27.968764	& 23.967	& 0.050	& 23.636	& 0.038	& 12685	& 21.311	& 0.011	& 0.187	& 1	& 23.570	& 0.027 \\
1	& 41130	& 52.934597	& -27.968940	& 27.583	& 0.221	& 25.409	& 0.039	& 12616	& 21.827	& 0.024	& 0.729	& 1	& 25.445	& 0.144 \\
2	& 41133	& 52.852864	& -27.968810	& 26.600	& 0.111	& 24.159	& 0.038	& 12556	& 22.492	& 0.021	& 0.616	& 1	& 24.164	& 0.049 \\
2	& 41133	& 52.852864	& -27.968810	& 26.600	& 0.111	& 24.159	& 0.038	& 12627	& 21.465	& 0.017	& 0.685	& 1	& 24.164	& 0.049 \\
1	& 41134	& 52.830309	& -27.968789	& 24.451	& 0.060	& 22.685	& 0.046	& 12683	& 19.897	& 0.004	& 0.535	& 1	& 22.707	& 0.016 \\
\hline
\end{tabular}
%\end{minipage}
\end{table*}

Next, we measure the FUV flux of those NUV-selected objects by applying DAOPHOT on the FUV images at the 
location of NUV-selected objects. Tests show that our FUV data are 80 \% complete down to FUV=26.8. Then, 
we perform a cross-correlation with COMBO 17 (Wolf et al. 2004) within a radius r=2 arcsecs. 
We find a counterpart in COMBO 17 for about 71\% of the NUV-selected sample. Table 1 provides the calibrated output from DAOPHOT (FUV and NUV) and  the aperture photometry within a 3-arcsec radius for the resulting 7309 objects inside the $0.263 ~deg^2$ in common with COMBO 17. \footnote{Some additional information such as the redshifts 
and the R-magnitudes extracted from COMBO 17 are also provided. However, the original COMBO 17 database 
should be consulted for complete information, using the listed COMBO 17 identification.} Some of the
sources have several possible counterparts (several NUV counterparts 
for one COMBO 17 source or vice-versa). An object with an observed color  $FUV-NUV \ge 2.0$ is 
classified as a member of our LBG sample if the redshift of the counterparts 
(sometimes several within 2 arcsecs) from COMBO 17 
is in the range $0.9 \le z \le 1.3$ (to be safe, we discard objects with variability flags larger than 8 and 
objects classified as "Star", "WD", "QSO"; "QSO (gal?)" and "Strange Objects"). At this point, we have 420 LBGs 
(2 with four COMBO 17 counterparts, 40 with two counterparts and 378 with one counterpart). We estimated that the total completeness of the NUV-selected sample is about 80\%. Correcting for this 80\% UV completeness, the observed surface density of NUV-selected galaxies (including LBGs) down to $NUV=26.2$ in the 0.26 sq. deg. is $6850 ~deg^{-2}$.

The 80 \% completeness is reached at FUV=26.8. Even though we still measure fluxes below this limiting magnitude, 
LBGs are missed for galaxies with $NUV \ge 24.8$ because FUV-NUV cannot be measured. However, if we restrict 
ourselves to $NUV < 24.8$ where we are able to estimate $FUV-NUV$ without the above restriction i.e. about 80\% completeness, the respective surface densities of NUV-selected objects and LBGs are $1766 ~deg^{-2}$ and $1320 ~deg^{-2}$ i.e. about 75\% of the NUV-selection are LBGs. A deeper analysis is postponed to Sect. 4.

Table 2 shows the completeness and the efficiency of the criterion $FUV-NUV \ge 2$ for identifying star-forming 
galaxies in the redshift range $0.9 \le z \le 1.3$. This method is very complete whenever the $FUV-NUV$ color 
is fully measurable i.e. when $NUV < 24.8$ as described above. A more efficient method over all redshifts 
would need to use color-color diagrams as suggested by the statistics presented by Adelberger et al. (2004). However, this is not the goal of this paper, where COMBO 17 redshifts are available as inputs to the sample selection. 

\begin{table}
\centering
%\begin{maxipage}{80mm}
\caption{The efficiency of our selection to identify $0.9 \le z \le 1.3$ galaxies through the detection of 
FUV dropout galaxies is 21.8 \% in the range $20.8 \le NUV \le 26.2$ but the completeness is very high at 83.3 \%.}
\begin{tabular}{@{}cccc@{}}
\hline
	$N_{UV-sel, all z}$	&	$N_{break, all z}$	&	$N_{LBG, 0.9 \le z \le 1.3}$	&	Efficiency	\\
\hline
				8206			&			2170			&										474	&	21.8 \%	\\
\hline
	$N_{UV-sel, 0.9 \le z \le 1.3}$	&	$N_{break, 0.9 \le z \le 1.3}$	&	$N_{LBG, 0.9 \le z \le 1.3}$	&	Completeness	\\
\hline
				1943			&			568				&										473	&	83.3 \%	\\
\hline
\end{tabular}
%\end{minipage}
\end{table}

Finally, we carried out cross-correlations to complete the wavelength coverage with the UBVRIJK ESO 
Imaging Survey (EIS), four IRAC bands and finally with MIPS at 24 $\mu m$ and $70 \mu m$. We stress that no 
additional selection effects enter into the final LBG sample since all the objects are kept in the following 
analysis with their own detections / non detections depending on the band. We have therefore, at the end of 
the cross-correlation process, the same number of UV objects as after the UV selection. Table 3 lists the best 
redshift as in Table 1, the UV $1800 \AA$ luminosity estimated by interpolating the observed NUV and 
U-band $\nu$ $f_{\nu}$ and the total IR luminosity evaluated from the $24 \mu m$ flux density and Chary \& 
Elbaz (2001) calibration. An interesting point is that 62 of the 420 LBGs (i.e. 15\%, but 17\% if we 
restrict our sample to $NUV < 24.8$ as above) have a $24 \mu m$ MIPS counterpart down to the MIPS/GTO limiting 
flux density of 83 $\mu Jy$ (all of them except one within r=2 arcsecs); we will call them Red LBGs 
(RLBGs hereafter). Huang et al. (2005) observed LBGs at $z \approx 3$ with Spitzer and, excluding objects 
with possible contamination from AGN, they also classified about 15\% as IR-luminous LBGs. Only 
two RLBGs are detected at $70 \mu m$ but none of them are detected at $160 \mu m$. The remaining LBGs, 
undetected in the MIPS image are called Blue LBGs (BLBGs). Note that this is a practical notation and it does 
not imply a physical difference so far. The observed surface densities are $1657 ~deg^{-2}$ for the much more 
numerous BLBGs, and $339 ~deg^{-2}$ for the RLBGs down to $NUV = 26.2$.

\begin{table*}
\centering
%\begin{maxipage}{80mm}
\caption{The LBG sample at $0.9 \le z \le 1.3$, as defined in the text, is listed here (466 rows). Redshifts are 
extracted from COMBO 17 and the redshift quality flag z$_{qual}$ corresponds to '1' if z is extracted from 
column "MC\_z" and '0' if z is extracted from column "MC\_z\_ml" in Wolf et al. (2004). No values '-99' are 
listed in the columns if the LBG is not detected in the U band (i.e. L$_{1800}$) or at $24 \mu m$ (L$_{dust}$) down to 83 $\mu Jy$, i.e. BLBGs. By definition, all objects are observed in NUV and should have a  L$_{1150}$ value. Objects with a L$_{dust}$ value are RLBGs. A$_{1600}$ is computed from the L$_{dust}$ / L$_{1600}$ ratio (L$_{1600}$ interpolated from the observed NUV and U bands).}

\begin{tabular}{@{}ccccccccc@{}}
\hline
id$_{UV}$  & RA$_{UV}$ & Dec$_{UV}$ & L$_{1150}$ & L$_{1800}$ & L$_{dust}$ &  z & z$_{qual}$ & A$_{1600}$ \\
\hline
\hline
46333	& 53.106075	& -27.918961	& 9.800	& 10.193	& 11.249	    & 0.900	& 1	&     2.092 \\
39913	& 53.252274	& -27.980711	& 9.540	& 10.006	& -99.000	& 0.900	& 0	& -99.000 \\
77655	& 52.931938	& -27.613210	& 10.054	& 10.185	& 11.008	    & 0.903	& 1	&     1.587 \\
50879	& 52.957910	& -27.876471	& 9.580	& 9.482	& -99.000	& 0.904	& 1	& -99.000 \\
53306	& 53.220624	& -27.853428	& 10.190	& 10.084	& -99.000	& 0.905	& 1	& -99.000 \\
\hline
\end{tabular}
%\end{minipage}
\end{table*}

\section{The Spectral Energy Distributions of Lyman Break Galaxies}

In this section, we analyse the median SEDs of RLBGs and BLBGs. In a previous paper (Burgarella et al. 2006b), 
we had stacked the 24 $\mu$m images of BLBGs in a first $z \sim 1$ LBG sample. Although we have improved our UV 
photometric measurements by using DAOPHOT, we consider that the stacked $24 \mu m$ flux densities for RLBGs 
given in Burgarella et al. (2006b) are valid because i) most of the LBGs in the first sample are included in 
the present one and ii) the stacking process is designed to reduce the influence of objects around the central 
detection (i.e. the LBGs) to enhance the signal-to-noise ratio of individually undetected BLBGs. The effect of 
the confusion is therefore reduced. Burgarella et al. (2006b) had estimated an average flux density of $13 \mu Jy$ 
at the observed wavelength of 24 $\mu$m for those BLBGs. Table 4 gives the absolute B magnitudes $M_B$ (from 
the observed I band) of BLBGs and RLBGs and their FUV absolute magnitudes. As order-of-magnitude comparisons, 
our LBGs are more luminous than Blue Compact Galaxies (e.g. $M_B<-18.5$ from Noeske et al. 2006); BLBGs are 
fainter than and RLBGs similar to LBGs at $z \sim 3$ (e.g. $<M_{1700} (BLBG)>=-21.0 \pm 1.0$ in Sawicki \& 
Thompson 2006).

\begin{table}
\centering
%\begin{maxipage}{80mm}
\caption{Average absolute magnitudes of the $z \sim 1$ LBG sample}

\begin{tabular}{@{}ccc@{}}
\hline
                 & $M_B$  & $M_{1800}$  \\
\hline
\hline
Blue LBGs  & $-21.0 \pm 1.0$ & $-20.3 \pm 1.0$ \\
Red LBGs   & $-22.4 \pm 1.0$ & $-20.9 \pm 1.1$ \\
 All LBGs   & $-21.0 \pm 1.1$ & $-20.3 \pm 1.1$ \\\hline
\end{tabular}
%\end{minipage}
\end{table}

Ideally, to relate our galaxies to each other it would be better to use the redshift of 
each galaxy to calculate K corrections and put them all in the same frame before building the template SED. 
However, since we cannot simply interpolate between observed measurements because of 
high contrast spectral features (for instance an 
interpolation between the observed 8 $\mu m$ and the observed 24 $\mu m$ bands would produce catastrophic 
results), we would need to use some fiducial SED for interpolation in redshift. 
Since there are no observed FUV-to-FIR SEDs built from a large sample of high redshift galaxies, 
we would have to rely on models. We 
prefer to avoid the resulting unknown uncertainties. Given our small redshift range, K-corrections 
should be relatively small. We plot in Figs. 1-3 the extent of the 
wavelength range (except in Fig. 3 hidden by symbols) due 
to the range in redshifts, showing that the errors are small from assuming an average redshift 
of $z = 1.1$ for all galaxies.

\subsection {Red Lyman Break Galaxies (RLBGs)}

We can study for the first time a large sample of LBGs in the wavelength range where the emission by dust is 
predominant. Moreover, two IR-bright LBGs in our sample have a significant piece of SED in the FIR with data points 
at $24 \mu m$ and $70 \mu m$. Interestingly enough, these two objects are consistent with being simultaneously 
ULIRGs (i.e. $L_{IR} \ge 10^{12} L_\odot$) and very blue (UV slope $\beta \sim -2$) UV luminous galaxies 
(UVLG) with moderate dust attenuation ($A_{FUV} \sim 1.8$). They will be described in a forthcoming paper 
(Burgarella et al. 2007). The other RLBGs are not detected at $70 \mu m$.

To build the UV-to-mm SED of RLBGs presented in Fig. 1, we proceed as follows: 
\begin{itemize}
\item{} We selected the RLBGs that are detected at all wavelengths from the observed NUV to $24 \mu m$ (i.e. 37 
objects). If we had computed the median SED of the full LBG sample, we would 
have derived a SED biased toward the brightest UV objects (and left undefined how to
treat the many upper limits at 24$\mu$m). By forcing the sample to have measured data points 
over the full wavelength range, we moderately bias the resulting SED toward red objects, 
which is consistent with our goal to characterize red LBGs. 
\item{} We normalize the median SED of two RLBGs detected up to $70 \mu m$ to the median SED of RLBGs in the 
observed IRAC $8 \mu m$ band. This normalization demonstrates that the $24 \mu m$ flux densities of 
these two RLBGs are very consistent with the $24 \mu m$ flux density of the median SED of the RLBGs. 
The values at other wavelengths are fairly consistent as well. 
The two SEDs are very likely extracted from the same parent RLBG population. Therefore, we make the assumption that 
the normalized $70 \mu m$ flux density of the two RLBGs detected at this wavelength can be used to extend the 
template SED of RLBGs. However, we must remember that the scatter in the two $70 \mu m$ data 
points is very likely not representative of a larger sample. 
\item{} We normalized the SED of HR 10, a dusty and luminous ($L_{TIR} \sim 9 \times 10^{12} L_\odot$) 
starburst at z=1.44 (Stern et al. 2006) to the observed IRAC 3.6$\mu$m band where the output should be dominated
by stellar photospheres ($\lambda_{rest} \sim 1.8\mu m$). 
It is interesting that the SED matches very well that of the median 
RLBG SED for rest-frame wavelengths in the range $1.8 \mu m$ - $70 \mu m$ (Fig. 1). Again, it seems that we 
are dealing with the same kind of objects (e.g. Elbaz et al. 2002). Therefore, we tentatively use the SED of 
HR 10 to extend our median SED to wavelengths longer than $70 \mu m$ up to 1mm. 
\end{itemize}

\noindent
The resulting RLBG template is listed in Table 5 and plotted in Fig. 1. The 
RLBG SED does not exhibit a power-law shaped continuum characteristic of AGN-dominated emission. 
Indeed, Ivison et al. (2004) showed that strong AGNs have nearly a constant slope from 2 to 10 $\mu m$ while 
a distinct minimum in the 3 to 4 $\mu m$ range is observed for starburst-dominated galaxies, and
this property is being used to identify AGN through IRAC colors (e.g., Lacy et al. 2004;
Stern et al. 2005; Alonso-Herrero et al. 2006). A key diagnostic 
for distinguishing AGN-dominated and starburst-dominated galaxies is a change of slope  between the 
stellar and dust continua at  3-6 $\mu m$, which is one of the most noticeable features in the SEDs of our RLBGs.

However, even though HR 10 bears strong similarities in the IR to our RLBG population, its SED is very faint 
in the optical and UV and we cannot simply state that RLBGs are HR 10 - like objects. The RLBGs might represent
a type of galaxy at $z \approx 1$ that could be considered to be a missing link between the two high redshift 
populations of galaxies: the blue and almost dust-free LBGs and the IR/sub-mm bright LIRGs/ULIRGs. 
We might be observing complex objects 
where one component is emitting in the UV and the other in the IR. A related object might be VV114 (Le Floc'h et al. 
2002),  which is also considered to be a local counterpart of distant LBGs. Do 
we see different classes of unrelated objects or is there an evolution from one class to 
the other where RLBGs represent a temporary phase? We do not have enough information, yet, 
to decide and a morphological analysis of the full sample is under way from HST imaging to provide clues 
toward or against this hypothesis.

To understand this median SED, we model it using PEGASE (Fioc \& Rocca-Volmerange 1997) in the UV + 
Optical + NIR (UVONIR)  and several modified blackbodies in the FIR and sub-mm ranges. In the UVONIR, the best 
match is for a 500 Myr-old model (exponentially decaying SFR $\Psi(t) = exp (-t/\tau)$ with $\tau = $ 7 Gyr, 
1/10 solar metallicity and a Salpeter IMF). To redden the model, we assume an attenuation law with 
$A_{FUV} \propto (\lambda / \lambda_{FUV})^n$ where $A_{FUV}$ is the FUV dust attenuation. Since all RLBGs 
have $A_{FUV}$ estimated from IR data, we used their mean $A_{FUV} = 2.45$ as a constraint. The slope, $n$, 
is a free parameter and $n = -0.7$ provides the best match to the observed SED. The UVONIR SED model is 
normalised to the observed R-band photometric point. 

In the MIR/FIR, we use four modified blackbodies 
with an emissivity index set to $\epsilon = 1.5$ (Klaas et al. 1997).
These blackbodies are not meant to reproduce the detailed emission spectrum in the PAH 
region (i.e. below about $12 \mu m$, e.g. Marcillac et al. 2006, Smith et al. 2006) but only the underlying 
continuum. As a consequence, dust luminosities estimated from this approach are, strictly speaking, only 
lower limits.  The temperatures of the four blackbodies are T = 300 K, 120 K, 45 K and 20 K. The normalization 
of the hottest one is fixed by setting the sum of the UVONIR model plus the 300 K model to the observed 
IRAC point at 8.0 $\mu m$. The warm blackbody is constrained by the 24$ \mu m$ MIPS point. The cool 
blackbody is fixed by the observation of the two RLBGs detected at 70$\mu m$ and finally the cold blackbody 
by the 350 $\mu m$ measurement of HR 10 (see above). The total IR luminosity from 1 to 1000 $\mu m$ amounts to 
$Log (L_{TIR}) = 11.5$ from the median RLBG SED. These RLBGs are therefore LIRGs on average. The UV 
luminosity integrated from 0.1 $\mu m$ to 1.0 $\mu m$ amounts to $L_{FUV} = 10.9$ but to estimate the 
$L_{TIR} / L_{FUV}$ ratio, we estimate $\lambda.f_\lambda (\lambda=1800 \AA) =  10.2$. These objects 
are therefore not UV luminous galaxies (UVLGs) on average. The resulting value of $Log(L_{TIR}) / L_{FUV} = 1.3$ 
converts into $A_{FUV} = 2.5$ using the calibration of Burgarella et al. (2006a). This value is in excellent agreement 
with the dust attenuation estimated from the fit.

\subsection {Blue Lyman Break Galaxies (BLBGs)}

To build the UV-to-mm SED of the BLBGs presented in Fig. 2, we proceed as follows: 

\begin{itemize}
\item{} The median SED of BLBGs is computed for BLBGs detected at all wavelengths from the observed NUV to 
the IRAC $8\mu m$ band (i.e. 68 objects). 
\item{} The $24 \mu m$ point is the result from the stacking carried out in Burgarella et al. (2006b).
\item{} The median SED of RLBGs presented above is superimposed onto the median SED of BLBGs after normalizing 
it to the observed IRAC $8 \mu m$ band, where stellar emission is still predominent. The match is fairly good 
from the IRAC 3.6$\mu$m band to $24\mu m$ and we use this normalized RLBG 
SED to extend the BLBG one, which means that we 
make the assumption that the temperatures of dust grains in BLBGs are the same as in RLBGs. Only the amount 
of dust attenuation differs.
\end{itemize}

\noindent
As for RLBGs, the median SED of the BLBGs does not show a power-law shaped continuum characteristic of 
AGN-dominated emission in the 3-4 $\mu m$ range.  As we showed for the RLBGs, the BLBGs should 
be starburst-dominated galaxies.

The UVONIR part of the SED is fitted again by a model from Fioc \& Rocca-Volmerange (1997). The best match 
is for a 250-Myr model with the same global characteristics as the RLBGs except for the amount of dust attenuation. 
For the BLBGs, we do not have any individual measurements of $L_{dust}$. A constraint can be set by the upper 
limit at 83 $\mu Jy$ flux density at 24 $\mu m$, which can be converted into an upper limit of $A_{FUV} = 2.1$. 
The best matching model corresponds to $A_{FUV} = 1.8$ with the same slope $n = -0.7$ as for the RLBGs. 
We conclude that the stellar populations in the BLBGs are only slightly younger than in the 
RLBGs and that the main difference between the two types comes from 
the amount of dust attenuation. The total IR luminosity of the median BLBG SED from 1 to 1000 $\mu m$ 
amounts to $Log (L_{TIR}) 
= 10.9$. A median BLBG is therefore not a LIRG. The UV luminosity integrated 
from 0.1 $\mu m$ to 1.0 $\mu m$ amounts to $L_{FUV} = 10.6$. Again, to estimate $L_{dust}/L_{FUV}$, we 
estimate $\lambda$ $f_\lambda (\lambda=1800 \AA) =  10.1$, which means that BLBGs are not UVLGs. The resulting
value of $Log L_{dust} / L_{FUV} = 0.9$ converts to $A_{FUV} = 1.8$ which, again, is consistent 
with the above value and with Schiminovich et al. (2005).

\subsection {Comparison with LBGs at $z \sim 3$}

Our two SEDs for BLBGs and RLBGs are compared in Fig. 3 to an average SED estimated for LBGs at $z \sim 3$ 
in F\"orster-Schreiber et al. (2004) and normalised at 2750 $\AA$. The grey cloud represents the standard 
deviation in F\"orster-Schreiber et al.'s LBG sample. Although marginally consistent with the latter, the 
SED of the RLBGs is too red in the UV and has an excess in the red part of the SED. Older stellar populations 
and/or more likely larger dust attenuation presumably are the origin of this difference. On the other hand, 
the SED of BLBGs has strong similarities to $z \sim 3$ LBGs: all the data points appear to be 
consistent, within the uncertainty, with the hypothesis that the two populations are extracted from the same 
sample. More data are needed in the rest-frame NIR and in the MIR/FIR to provide information on the evolved stellar 
population and on the amount of dust to reach a definite conclusion. However, the present data and 
the results of Huang et al. (2005) suggest that we might observe LBGs at $z \sim 3$ with the same SED 
characteristics (shape and distribution) as LBGs in the $z \sim 1$ universe. Moreover, the LBG selection 
seems to favour starbursting galaxies with low dust attenuation whatever the redshift, that is BLBGs as 
defined in our sample. Nevertheless dustier LBGs (i.e. RLBGs) exist at $z \sim 1$ and $z \sim 3$ with a 
higher total SFR than BLBGs. However, we saw in Table 4 that LBGs at $z \sim 3$ are brighter than BLBGs 
but have magnitudes similar to RLBGs. This means that a simple identification of BLBGs with high redshift 
LBGs is not strictly valid.

\begin{table*}
%\begin{maxipage}{80mm}
\caption{Measured median SED of 37 RLBGs and 68 BLBGs $\nu$ $f_{\nu}$ (erg.cm$^{-2}$.s$^{-1}$). 
For RLBGs, the 70 $\mu m$ data are extracted for two RLBGs only. The 24 $\mu m$ data for BLBGs have been 
estimated by stacking BLBGs (Burgarella et al. 2006b).}

\begin{tabular}{@{}cccccccc@{}}
\hline
$\lambda_{obs}~(\mu m)$  & 0.23 & 0.36 & 0.44 & 0.55 & 0.71 & 0.97 & 1.25 \\
\hline
\hline
$\nu$ $f_{\nu}$ (RLBG)  &    8.31E-15 &   1.47E-14 &   1.30E-14 &   1.27E-14 &   1.13E-14 &   1.63E-14 &  1.84E-14  \\
\hline
1$^{st}$ quartile (RLBG)  &    6.30E-15 &   7.68E-15 &   7.68E-15 &   7.34E-15 &   8.74E-15 &   1.35E-14 &   1.64E-15 \\
\hline
3$^{rd}$ quartile (RLBG)  &    1.28E-14 &   2.11E-14 &   1.81E-14 &   1.71E-14 &   1.69E-14 &   2.53E-14 &   2.50E-15 \\
\hline
$\nu$ $f_{\nu}$ (BLBG)  &    7.16E-15 &   8.41E-15 &   7.25E-15 &   5.80E-15 &   5.27E-15 &   6.14E-15 &   5.05E-15 \\
\hline
1$^{st}$ quartile (BLBG)  &    4.84E-15 &   5.66E-15 &   5.22E-15 &   4.29E-15 &   3.80E-15 &   3.91E-15 &   1.64E-15 \\
\hline
3$^{rd}$ quartile (BLBG)  &    8.31E-15 &   1.13E-14 &   9.57E-15 &   8.15E-15 &   7.56E-15 &   8.85E-15 &   8.13E-15 \\
\hline
\hline
$\lambda_{obs}~(\mu m)$  & 2.2 & 3.6 & 4.5 & 5.8 & 8.0 & 24 & 70 \\
\hline
$\nu$ $f_{\nu}$ (RLBG)  &   1.18E-14 &   1.11E-14 &   6.15E-15 &   3.71E-15 &   2.05E-15 &   1.09E-14 &   9.17E-14 \\
\hline
1$^{st}$ quartile (RLBG)  &    2.32E-15 &   6.98E-15 &   3.77E-15 &   2.13E-15 &   9.88E-16 &   5.86E-15 &   2.54E-13 \\
\hline
3$^{rd}$ quartile (RLBG)  &    2.15E-14 &   1.74E-14 &   1.04E-14 &   5.80E-15 &   3.62E-15 &   1.90E-14 &   2.83E-13 \\
\hline
$\nu$ $f_{\nu}$ (BLBG)  &    2.32E-15 &   2.56E-15 &   1.46E-15 &   8.23E-16 &   4.55E-16 &   1.63E-15 & - \\
\hline
1$^{st}$ quartile (BLBG)  &    2.32E-15 &   1.57E-15 &   9.85E-16 &   4.86E-16 &   2.13E-16 &   - &   - \\
\hline
3$^{rd}$ quartile (BLBG)  &    6.18E-15 &   3.92E-15 &   2.47E-15 &   1.36E-15 &   7.49E-16 &   - &   - \\ 
\hline
\hline
\end{tabular}
%\end{minipage}
\end{table*}

\begin{figure*}
%\vspace{2pt}
\epsfxsize=16truecm\epsfbox{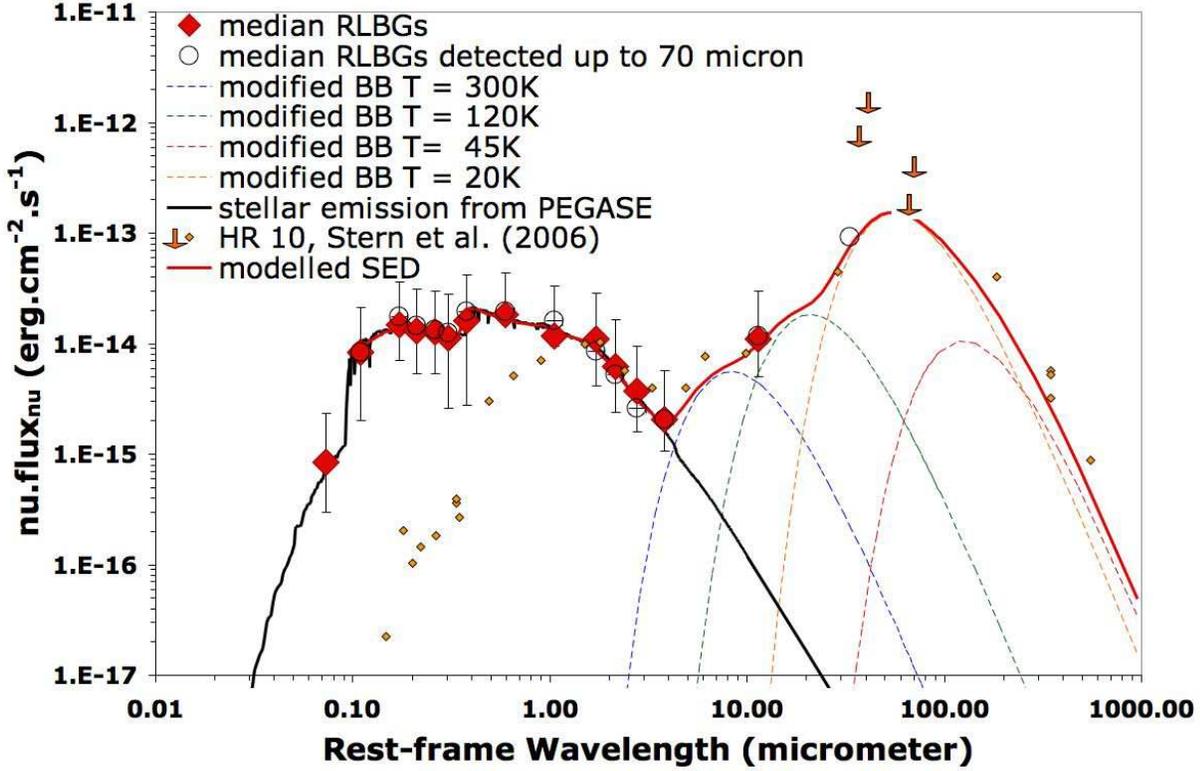}
\caption{\label{fig1} The average SED of RLBGs is modelled using a reddened PEGASE model in the UVONIR and 
modified blackbodies in the FIR and submm ranges. Good agreement with the SED of HR10 exists in the IR, 
showing that we see very dusty galaxies. However, unlike HR10, RLBGs are also detected in the UV.}
\end{figure*}

\begin{figure*}
%\vspace{2pt}
\epsfxsize=16truecm\epsfbox{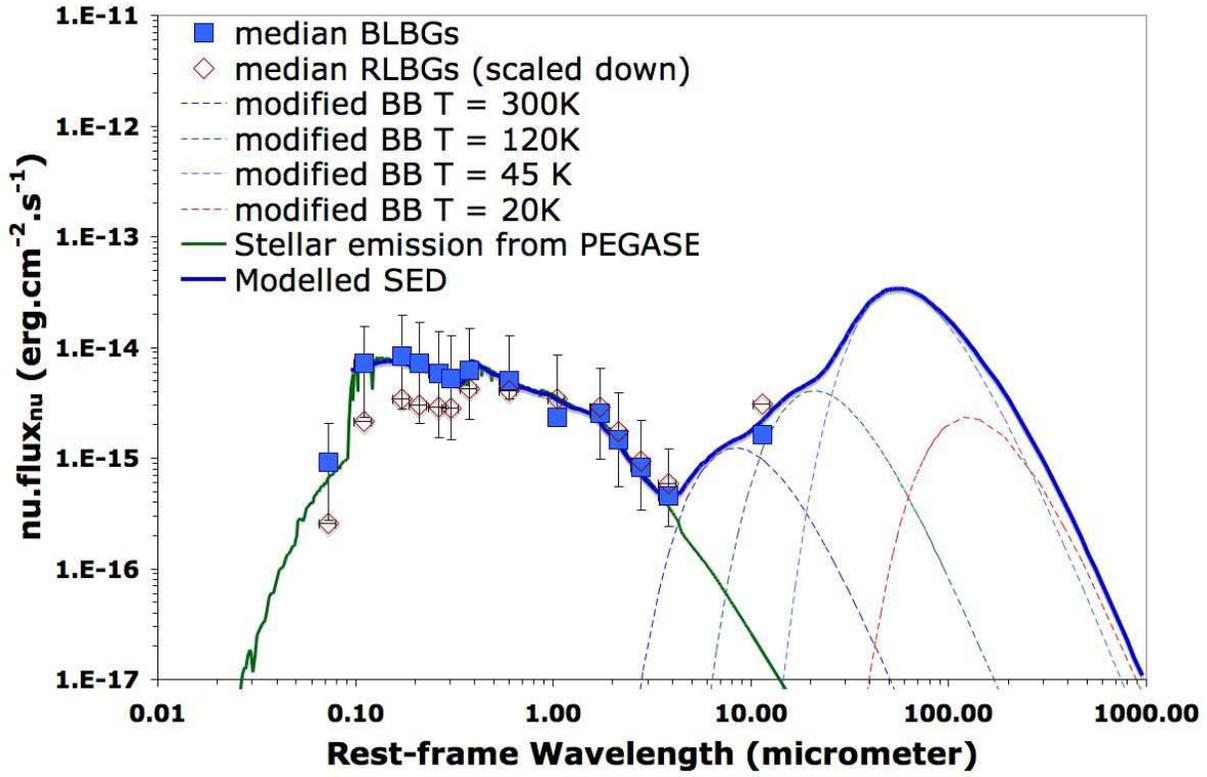}
\caption{\label{fig2} The average SEDs of BLBGs looks similar to that of RLBGs but we have to assume a lower 
dust content to explain the UVONIR SED. This is in agreement with the non direct detection at $24 \mu m$. 
The value at this wavelength is provided by stacking.}
\end{figure*}

\begin{figure*}
%\vspace{2pt}
\epsfxsize=16truecm\epsfbox{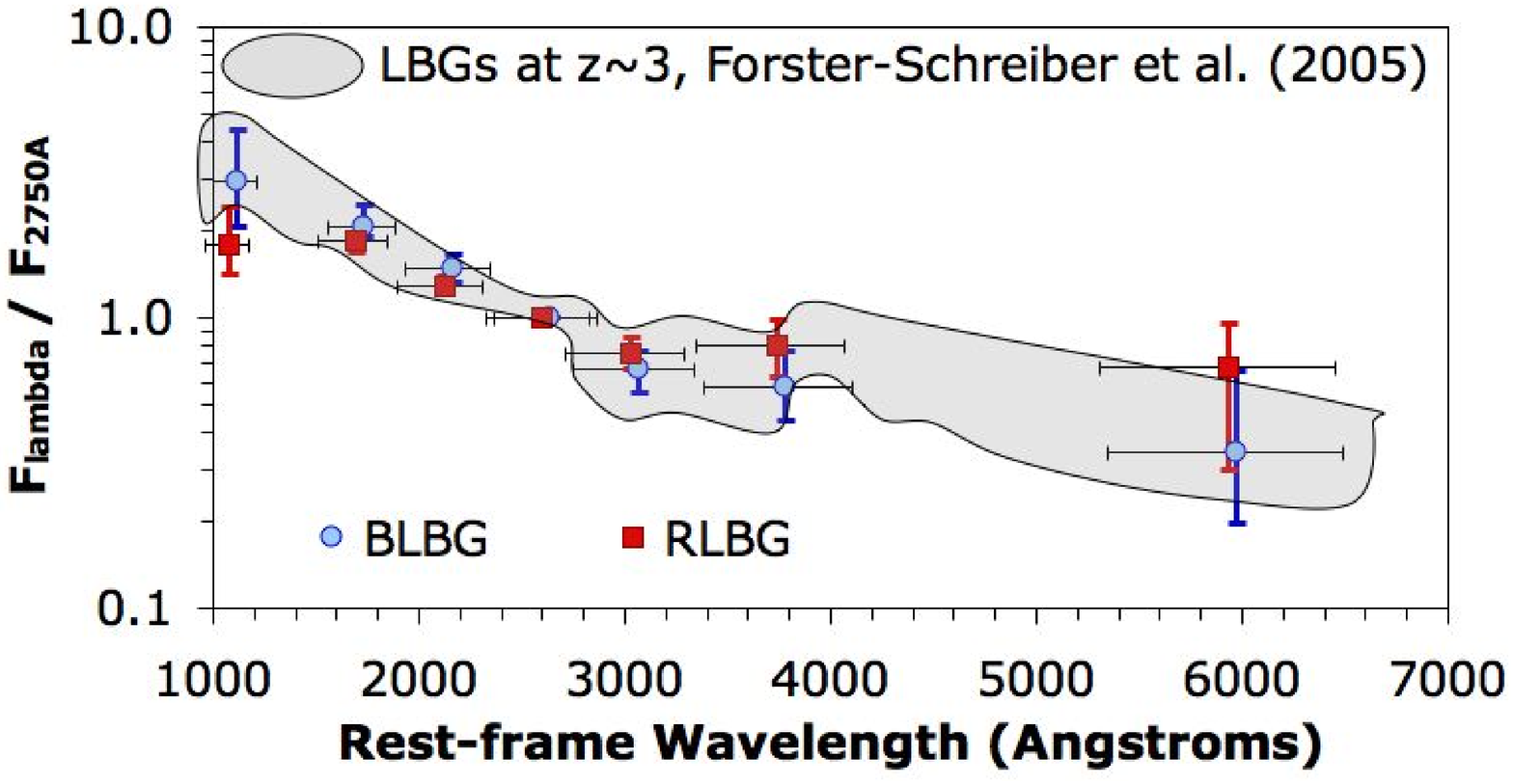}
\caption{\label{fig3} The UV/optical SEDs of the two classes of LBGs are noted as in Figs. 1 and 2. They 
are compared to the average SED of LBGs at $z \sim 3$ by F\"orster-Schreiber et al. (2004) located in the 
grey area. The comparison suggests that the latter are more similar to BLBGs than to RLBGs on average.}
\end{figure*}

\section{The Ultraviolet Luminosity Function of Lyman Break Galaxies}

The LBG selection process preferentially detects star-forming galaxies with low dust 
attenuation (see also, e.g., Steidel et al. 1996). This is generally true for UV-selected 
galaxies whereas IR-selected galaxies tend to be star-forming 
galaxies with a high $L_{dust}/L_{FUV}$ ratio, that is, high dust attenuation (Iglesias-Paramo et al. 2006, 
Buat et al. 2006, Burgarella et al. 2006a). Xu et al. (2006) suggest that these differences are mainly 
caused by selection effects: both UV-selected and IR-selected galaxies would be extracted from the same 
population of star-forming galaxies. Buat et al. (2006) reach a somewhat similar conclusion. What 
can we say about high redshift LBGs? 

LBGs are selected through colours. Therefore, strictly speaking they do not form a genuine UV-selected 
sample. Their spectral break is caused by the absorption of UV photons with $\lambda < 912 \AA$ in the outer atmospheres 
of massive stars, in interstellar H I gas, and in intervening H I gas along the line of sight. However, in 
the case of star-forming galaxies, the flux redward of the Lyman break (in the rest frame UV) is high, and high 
redshift LBGs are often assumed to be similar to a UV selected sample (Giavalisco 2002). High redshift LBGs are 
therefore used to estimate the UV star formation density. However, the intergalactic medium has little 
effect in producing the Lyman break in the SED of LBGs at $z \sim 1$ and our objects are true Lyman break 
galaxies for which the break is produced by material inside the observed galaxies. We need therefore to 
check whether our LBG selection is equivalent to a UV selection before we can  put local and high 
redshift LBG studies on the same ground to compare them.

To address these questions, we need to compare the luminosity function (LF) of a UV-selected sample to 
the LF of a LBG sample. We address this point at $z \sim  1$. Arnouts et al. (2005) computed the LF of 
UV-selected galaxies at 150nm and $z \sim  1$. We evaluate the LF of our NUV-selected sample using the new DAOPHOT 
photometry and the $1/V_{max}$ and $C-$ methods, to which we applied a K-correction to get LFs in the same 
wavelength range as for the higher redshift studies (e.g. Madau et al. 1996, Steidel et al. 1999, Sawicki \& 
Thompson 2006). Both LFs at $z \sim 1$ are very consistent (Fig. 4). We also compute the Far-UV (FUV) 
LF of our LBG sample (using $1/V_{max}$ and $C-$). To select the z $\sim$ 1 LBG sample, we use a color selection 
$FUV-NUV > 2$. This means that there is a secondary selection in addition to the primary FUV selection 
on the sample. This secondary selection is properly addressed by the Kaplan-Meier estimator. By this 
method, we can utilize the information contained in the upper limits (see a more detailed description 
in Xu et al. 2006), and correctly estimate the LBG luminosity function. We compare it to the two FUV 
LFs in Fig. 4. 

The UV luminosity density $\rho_{FUV}(1800A)$ from Schiminovich et al. (2005) is 
$\rho^{Sch}_{FUV}(1800\AA) = 9.3\times 10^7 h [L_\odot Mpc^{-3}]$. If we restrict Arnouts' LF to our 
sample range, we obtain $\rho^{Sch-restr}_{FUV}(1800\AA) = 5.1\times 10^7 h [L_\odot Mpc^{-3}]$ 
while our NUV-sel LF gives 

$$\rho_{FUV}(1800\AA) = 4.0 \pm 0.6 \times 10^7 h [L_\odot Mpc^{-3} ].$$

\noindent
The difference amounts to $\sim$ 20\%. For LBGs, 

$$\rho_{FUV}^{LBG}(1800\AA) = 3.4 \pm 0.7 \times 10^7 h [L_\odot Mpc^{-3}]$$

\noindent
which represents 85\% of the total $\rho_{FUV}(1800\AA)$ of NUV-selected galaxies in the considered 
luminosity range. The given uncertainties are 20\%, so the two values are consistent. We can conclude 
that at $z \sim 1$, our LBG selection (which might be different from a higher redshift LBG selection 
i.e. without screening objects with high $A_{FUV}$) is similar to a UV selection. 

Since most of our LBGs (83\%) belong to the BLBG class, the difference between the LFs of BLBGs and LBGs  
is small. As expected, the LF of RLBGs ($\sim $17~\% of the sample) falls well below. While the shape 
above $Log (L_{FUV}/{L_\odot}) = 10.3$ (i.e. UVLGs) of the RLBG LF is the similar to that of all LBGs, 
the RLBG LF seems to flatten at low luminosity. The contributions from blue and red LBGs to the LBG UV 
luminosity density are as follows:

$$\rho_{FUV}^{BLBG}(1800\AA) = 2.9 \pm 0.9 \times 10^7 h [L_\odot Mpc^{-3}]$$

\noindent
i.e. 72.5\% of the FUV flux density, and

$$\rho_{FUV}^{RLBG}(1800\AA) = 2.7 \pm 1.1 \times 10^6 h [L_\odot Mpc^{-3}]$$

\noindent
i.e. 7\% of the FUV flux density.

\begin{figure}
%\vspace{2pt}
\epsfxsize=8truecm\epsfbox{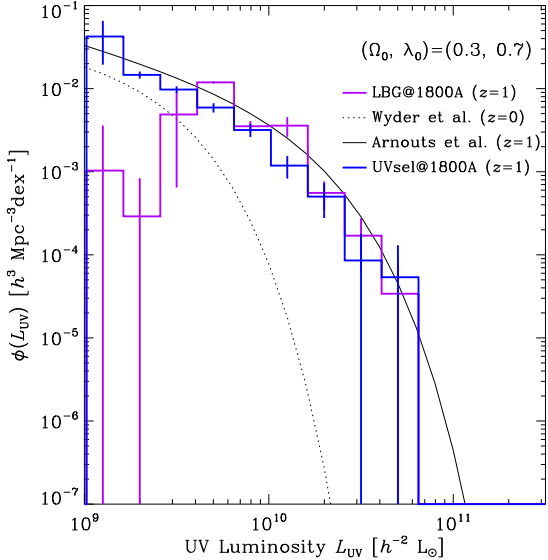}
\caption{The luminosity function of all LBGs estimated from the $1/V_{max}$ method and confirmed by 
the $C-$ method. At $Log (L_{FUV}) > 2 \times 10^{10}$ i.e. UVLGs, this luminosity function is almost 
the same as the NUV luminosity function of UV-selected galaxies at $z=1$ from our own measurements 
and from Arnouts et al. (2005).\label{fig4}}
\end{figure}

%\begin{figure}
%\vspace{2pt}
%\epsfxsize=8truecm\epsfbox{BRLBGLFFig7.eps}
%\caption{The green solid line is the luminosity function of all LBGs estimated from the V$_{max}$ 
%method. The luminosity function of BLBGs presents the same shape than that of all LBGs, while the 
%luminosity function of RLBGs is lower and might present a flattening at low luminosity.\label{fig7}}
%\end{figure}

\section{Contributions of UV- and IR-deduced SFR to the total Star Formation Density}

Assuming a 250-500 Myr exponentially decaying star formation rate from PEGASE, in agreement with the 
SED fitting in the previous section, we can estimate $SFR_{1150}$ for the UV-selected sample and for LBGs:

$$SFR_{1150} = \nu  L_\nu ~[erg ~ s^{-1}] / ~1.8 \times 10^{-30} M_\odot ~ yr^{-1} ~ Mpc^3$$

\noindent
In determining this value, we used our 80~\% complete NUV measurements, which apply to a 
rest-frame wavelength of $1150 \AA$. This wavelength corresponds to the observed V-band at $z \sim 3 - 4$. 
Since we will make comparisons within our own dataset, adopting this wavelength reduces the uncertainties 
due to K-corrections.

Using the observed $24 \mu m$ flux density we can estimate the luminosity from emission by dust,
$L_{TIR}$, from the Chary \& Elbaz (2001) calibration. However, it is necessary to evaluate how 
the aromatic bands could affect the estimated infrared luminosity. 
Based on analysis of a sample of local galaxies and assuming several galaxy SEDs, Dale et al. (2005) 
show the evolution of the ratio $f_{TIR}$ / $\nu$ $f_{\nu} (24 \mu m)$ with the redshift. In our redshift range, 
their figure suggests that the ratio can change by a factor of 2 - 3 for their galaxy sample. However, 
our galaxies are all luminous starburst galaxies. Smith et al. (2007) estimated 
that the ratio of the PAH-to-TIR luminosities for starburst galaxies 
$L(PAH) / L_{TIR} \sim  0.11 \pm 0.04$.  Marcillac
et al. (2006) consider the scatter of the mid-IR indicators against each other and the radio (which should
be a measure of the far infrared) and demonstrate that, for the best fitting templates, the scatter
is only of order 30-40\% rms. 
Using the relative band strengths in Smith et al., 
we can estimate that only $\sim 25 \%$ of the flux density in the MIPS $24 \mu m$ band 
is due to aromatic features in our redshift range. Since we apply the rest-frame $12 \mu m$ calibration 
from Chary \& Elbaz (2001) to estimate $L_{TIR}$ from $\nu$ $f_{\nu} (24 \mu m)$, we include
the aromatic contribution to first order. Therefore, the relative contribution 
of the aromatic features to the $24 \mu m$ signal is not a major issue. 
As we show later, the ratio for our two galaxies 
detected at 70$\mu$m,  $f_{\nu} (70 \mu m) / f_{\nu} (24 \mu m) \approx 15$, is consistent with warm dust 
temperatures and of the same order as typical starburst galaxies such as Mark 33 in Dale et al. (2005). 
We therefore expect only a modest dispersion in the calibration from 
$\nu$ $f_{\nu} (24 \mu m)$ to $L_{TIR}$ (e.g. Papovich et al. 2002). 

More crucial is the possibility that aromatic features 
move in and out of the $24\mu m$ MIPS band in the $0.9 \le z \le 1.3$ redshift range, creating
a variable calibration. The main features of interest 
are at 11.3 $\mu m$, 12.0 $\mu m$ and 12.6 $\mu m$. Any variability due to the first and second features is 
negligible since they remain within the MIPS $24 \mu m$ band over almost all our redshift range. However, the 
12.6 $\mu m$ feature (plus the 12.8 $\mu m$ [NeII] line) moves out of the filter at $z \sim 1.1$. To check 
whether this produces a calibration change (namely a decrease of the estimated luminosity above $z = 1.1$), we divide 
our sample into two ranges below and above $z = 1.1$. The ratio of the $24 \mu m$ flux densities is about 
unity in the two redshift ranges. This implies an increase of $SFR_{TIR}$ by a factor of $\sim 1.5$ with 
the redshift due to geometrical effects, which is consistent with our calculations. 
The average dust attenuations we calculate in the two adjacent 
redshift ranges are constant. We conclude that the derived behavior of the galaxies
has no discontinuity at z $\sim$ 1.1 and that there is no strong variation due to the aromatic feature
moving out of the 24$\mu$m band. Therefore, we will assume that we can 
trust our estimated $L_{TIR}$. 

We apply Kennicutt's (1998) calibration of $L_{TIR}$ into $SFR_{TIR}$, 
which is valid for starbursts:

$$SFR_{TIR} = 4.5 \times 10^{-44} L_{TIR} ~[erg.s^{-1}] M_\odot.yr^{-1}.Mpc^3,$$

As already noticed in Sect. 2, our observed LBG sample is not complete below 
$NUV=24.8$ because of the FUV 80\% limiting magnitude at $FUV=26.8$. To estimate the total SFRs and surface 
densities down to $NUV=26.2$, we assume that the characteristic percentages evaluated where we are complete 
can be applied below $NUV = 24.8$. Namely, that means that we assume, at $NUV<24.8$ and $NUV\ge 24.8$, the 
same percentages of LBGs in the UV-selected sample, the same percentages of BLBGs and RLBGs in the LBG sample 
and identical contributions to SFRs for the above classes. In other words, we make the assumption that the 
instrumental threshold is not related to a physical one and that no important trends exist as a function of 
the FUV luminosity $L_{FUV}$. This is very likely not true over a wide range of $L_{FUV}$ but the two $L_{FUV}$ 
ranges (median, standard deviation) above and below $NUV=24.8$ are similar at 
$1.4 \times 10^{10} \pm 1.1 \times 10^{10} L_\odot$ and  $0.5 \times 10^{10} \pm 0.2 \times 10^{10} L_\odot$,
respectively.

Although they are more numerous, BLBGs undetected at $24 \mu m$ do not contribute much to the $SFR_{dust}$ 
budget but only to $SFR_{1150}$ one; they represent 23\% of the LBG $SFR_{TOT}$. RLBGs provide most of the LBG 
$SFR_{TOT}$ (77\%) deduced from both the UV and infrared emission. However the surface densities of BLBGs and 
RLBGs follow an opposite trend: 83 \% are BLBGs and 17 \% are RLBGs. Most LBGs are therefore blue and 
contribute to a small percentage of the LBG $SFR_{TOT}$ while only a small proportion of dusty LBGs 
contribute the bulk of the LBG $SFR_{TOT}$.

At $1150 \AA$, LBGs furnish 69\% of $SFR_{1150}$. The ratio $SFR_{TIR}/SFR_{1150} = 24.6$ is larger than 
the ratio of SFR densities $\rho_{TIR}/\rho_{UV}=5.2$ found at $\lambda=1500\AA$ and in the redshift range 
$0.8 \le z \le 1.2$ by Takeuchi, Buat \& Burgarella (2005). However, the data in Table 6 are integrated over 
the total IR luminosity range $L_{TIR} > 10^{11} L_\odot$ and $L_{1150} > 5 \times 10^{9} L_\odot$. In addition 
to the wavelength difference, this might explain the different ratio. Indeed, if we sum up our TIR SFR 
density $\rho_{TIR} = 0.11 M_\odot ~ yr^{-1}~ Mpc^{-3}$.
For comparison, we can estimate $SFR_{TIR}$ for the IR-selected sample of LIRGs and ULIRGs defined by 
Le Floc'h et al. (2005) in the same CDFS 
field down to $83 \mu Jy$. The result is very consistent: $\rho_{TIR} = 0.12 M_\odot ~ yr^{-1} ~ Mpc^{-3}$. 
On the other hand, from the FUV luminosity density provided by Schiminovich et al. (2005) for 
$L _{FUV}> L_{min} = 0.2 L_{*,z \sim3 }$, we can estimate $SFR_{1500} = 0.0045 M_\odot.yr^{-1} ~ Mpc^{-3}$ 
in the same FUV luminosity range, which is very consistent with our sum $SFR_{1150} = 0.0046 
M_\odot ~ yr^{-1} ~ Mpc^{-3}$ in the UV luminosity $L_{1150} > L_{min} = 0.1 L_{*,z \sim3 }$ estimated 
from $NUV_{lim} = 26.2$. From the SFR densities provided by Le Floc'h et al. (2005) and Schminovich 
et al. (2005), we can compute: $\rho_{TIR}/\rho_{FUV}=26.6$, which is very close to our 
ratio. A more complete analysis requires a more sophisticated analysis using bivariate LFs that will 
be performed in a forthcoming paper (Takeuchi et al. 2007).

\begin{table*}
%\begin{maxipage}{80mm}
\caption{Contribution of $1150 \AA$ UV SFR ($SFR_{1150}$) and dust SFR ($SFR_{dust}$) of LBGs and other 
non-LBG objects to the total SFR. SFR$_{1150}$ is computed from observed GALEX NUV magnitudes (i.e. rest-frame 
FUV) whereas $SFR_{dust}$ is computed from observed Spitzer/MIPS $24 \mu m$ flux density, transformed 
into $L_{dust}$ using Chary \& Elbaz (2001) and into $SFR_{dust}$ using Kennicutt's (1998) relation.}

\begin{tabular}{@{}lccc@{}}
\hline
   & SFR$_{LBG}$ & SFR$_{not-LBG}$ & SFR$_{ALL}$ \\
   & (BLBGs; RLBGs) &  (UV; IR) &  \\
\hline
\hline
SFR$_{1150}$ ($M\odot.yr^{-1}.deg^{-2}$)    & 11086 (78 \%; 22 \%) &     4937; 0 &   16023 \\
SFR$_{IR}$ ($M\odot.yr^{-1}.deg^{-2}$)         & 27029 (0 \%; 100 \%) & 0; 365445 & 394474 \\
SFR$_{TOT}$ ($M\odot.yr^{-1}.deg^{-2}$)      &38115 (23 \%; 77 \%) &  372382 & 410497\\
Surface Density ($deg^{-2}$)                          & 6003 (83 \%; 17 \%)  & 1733;4312 & 6851; 5197 \\
\hline
\end{tabular}
%\end{minipage}
\end{table*}

%\begin{figure*}
%\vspace{2pt}
%\epsfxsize=12truecm\epsfbox{TABSFRFig5.eps}
%\caption{Contribution of $1150 \AA$ UV SFR ($SFR_{1150}$) and dust SFR ($SFR_{dust}$) of LBGs and other non-LBG objects to the total SFR. SFR$_{1150}$ is computed from the observed GALEX NUV magnitudes (i.e. rest-frame FUV. \label{fig5}}
%\end{figure*}

\section{Can we estimate total star formation rates of UV-selected galaxies from UV and/or IR ?}

Since the early works by Steidel et al. (1996) and Madau et al. (1996), astronomers have tried to 
evaluate the cosmic star formation rate and its evolution with redshift from UV-selected galaxies. 
However, a correction for the dust attenuation must be applied to the UV measurements to obtain the total 
star formation density. The most widely used method is based on the Meurer, Heckman \& Calzetti 
(1999) $IRX-\beta$ relation, whose validity is still debated (e.g. Goldader et al. 2002, Buat et al. 2005). 
Other methods have also been proposed (e.g. Kong et al. 2004, Burgarella et al. 2006a). 
Burgarella et al.'s (2006a) two-colour method, estimated for a sample of normal and local galaxies, does 
not seem to provide good dust attenuation estimates, which might indicate different dust attenuation laws 
for these local galaxies than for higher redshift star-forming LBGs. Burgarella et al. (2006b) presented 
a comparison of $SFR_{TOT}$ (i.e. FUV + dust) estimated from the FUV uncorrected for dust attenuation, 
the FUV corrected for dust attenuation using the $IRX-\beta$ method and from IR emission. In this paper, 
we update and complete this picture (Fig. 5). The main result is that IR-based SFRs provide the best estimate for 
$SFR_{TOT}$ of RLBGs, which are the only LBGs for which we have a direct estimate of the total SFRs. 
This was expected since RLBGs are dominated by their IR emission. It will be interesting, though, to extend 
the diagram to lower SFRs to follow the evolution. Iglesias-Paramo et al. (2006) performed a similar 
comparison at $z \sim 0$ and found that $SFR_{FUV}$ and $SFR_{dust}$ contribute about the same percentage 
to $SFR_{TOT} \approx 15 M_\odot ~ yr^{-1}$ which is not inconsistent with an extrapolation from our present data. 
On the other hand, the $IRX-\beta$ method seems to provide, on average, a correct estimate of $SFR_{TOT}$, 
even for IR-bright LBGs. However, the dispersion is much larger that for FIR-based SFRs.

\begin{figure*}
%\vspace{2pt}
\epsfxsize=16truecm\epsfbox{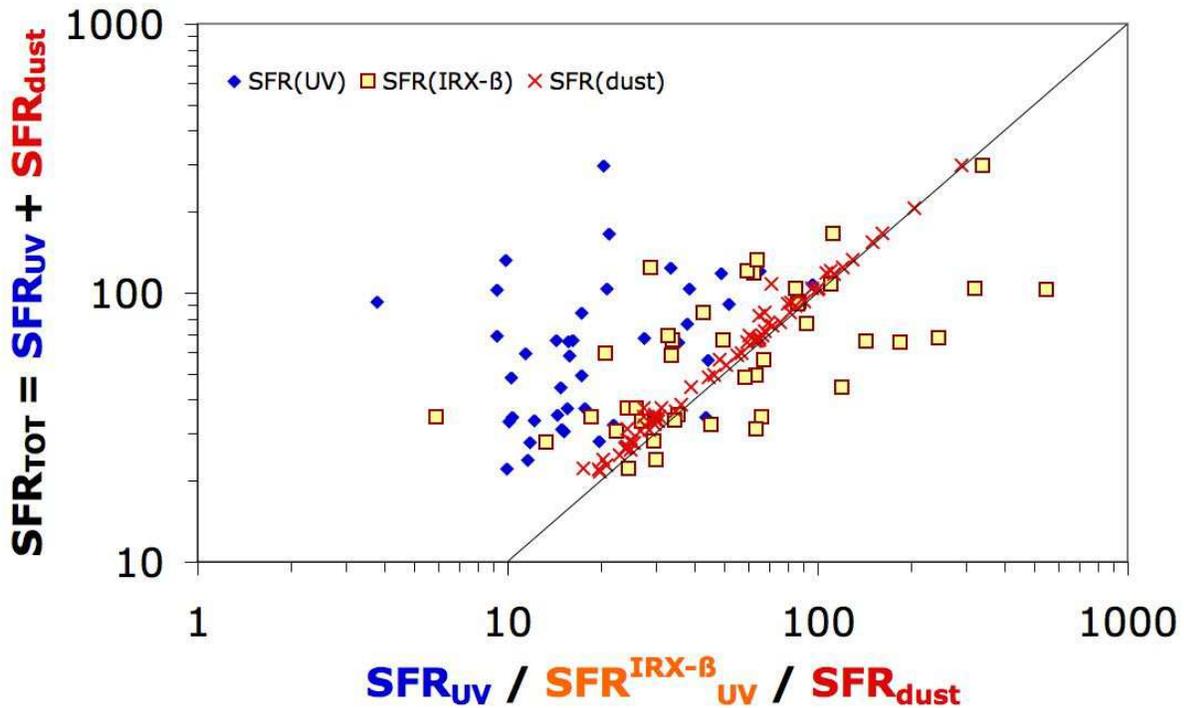}
\caption{For LBGs detected at $24 \mu m$ with Spitzer/MIPS, SFRs estimated from the dust emission are the best 
tracer of total SFRs. However, those objects are IR bright and this good correlation is expected. SFRs 
estimated from the UV without correcting for the dust attenuation underestimate the total SFRs. Finally, SFRs 
evaluated by applying the $IRX-\beta$ method give a rough order of magnitude for the total SFRs but the 
dispersion is much larger than for IR-based SFRs.\label{fig5}}
\end{figure*}

\section{The Evolution of the Total IR to Far UV luminosities of Lyman Break Galaxies}

Buat et al. (2007b) defined a sample of 190 LIRGs at $z \sim 0.7$ in the CDFS and compared it to a sample of 
120 LIRGs at $z \sim 0$ extracted from Buat et al. (2005). They found that the distribution of $L_{TIR}/L_{FUV}$ 
is different at the 95\% level, as a result of different average FUV dust attenuations: $<A_{FUV} (z=0)> = 3.8 \pm 0.1$ 
and $<A_{FUV} (z\sim 0.7)> = 3.4 \pm 0.1$. Since the luminosity function of LBGs is statistically consistent 
with the luminosity function of a UV selection, we can compare $L_{TIR}/L_{FUV}$ for 
our objects to the UV-selected sample at $z = 0$ of Buat et al. (2007a) and the BM/BX FUV-selected 
sample at $z \sim 2$ of Reddy et al. (2006). Unfortunately, Buat et al.'s sample does not 
extend far into the $10^{11} L_\odot$ regime where galaxies lie at $z \sim 1$ and $z \sim 2$, making a direct 
comparison impossible. However, Buat et al.'s FUV-selected and FIR-selected samples do not show any 
difference up to $Log (L_{TOT}) \approx 11$. Therefore, we adopt two values at $z = 0$ for the sake of a 
redshift comparison of the  ($L_{dust}/L_{FUV}$) ratio: first, we assume that we can extrapolate the FUV 
trend to higher $Log (L_{TOT})$ through a linear regression and second we use the FIR value. At 
$Log (L_{TOT}) = 11.5$ (i.e. a LIRG), the typical value of $L_{TIR}/L_{FUV}$ for the volume-corrected 
sample at $z = 0$ is $1.92 \pm 0.52$ for the FIR-selected galaxies and $1.53 \pm 0.36$ for the FUV-selected 
ones. The comparative values are $1.22 \pm 0.23$ at $z = 1$ and $0.84 \pm 0.22$ at $z = 2$ (Fig. 6). The corresponding FUV 
dust attenuations, using the $L_{TIR}/L_{FUV}$ to $A_{FUV}$ calibration provided by Burgarella et al. (2006a), 
are respectively $A_{FUV} = 3.9, 3.0, 2.4$ and 1.7.

We clearly observe a regular decrease of $L_{TIR}/L_{FUV}$ with redshift (Fig. 7). What is the evolution 
beyond $z = 2$? We need more data at all luminosities and, of course, a better coverage of the redshift scale. 
The evolution of this ratio will be a very powerful tool to bring us some constraint on the dust formation 
timescale and process in galaxies with, later on, some idea of where in the UV or FIR we will have to 
look for high redshift star formation. Ultimately, dust-free star formation might be observed at an unknown 
redshift depending on the time needed to form dust in star forming regions. We must stress, however, that 
this decrease is based on a UV-selected sample (which is generally associated with lower stellar masses 
than IR-selected galaxies). Buat et al. (2007) find a smaller effect for an 
IR-selected sample at $z \approx 0.7$, which might suggest a mass-dependent phenomenon.

\begin{figure*}
%\vspace{2pt}
\epsfxsize=16truecm\epsfbox{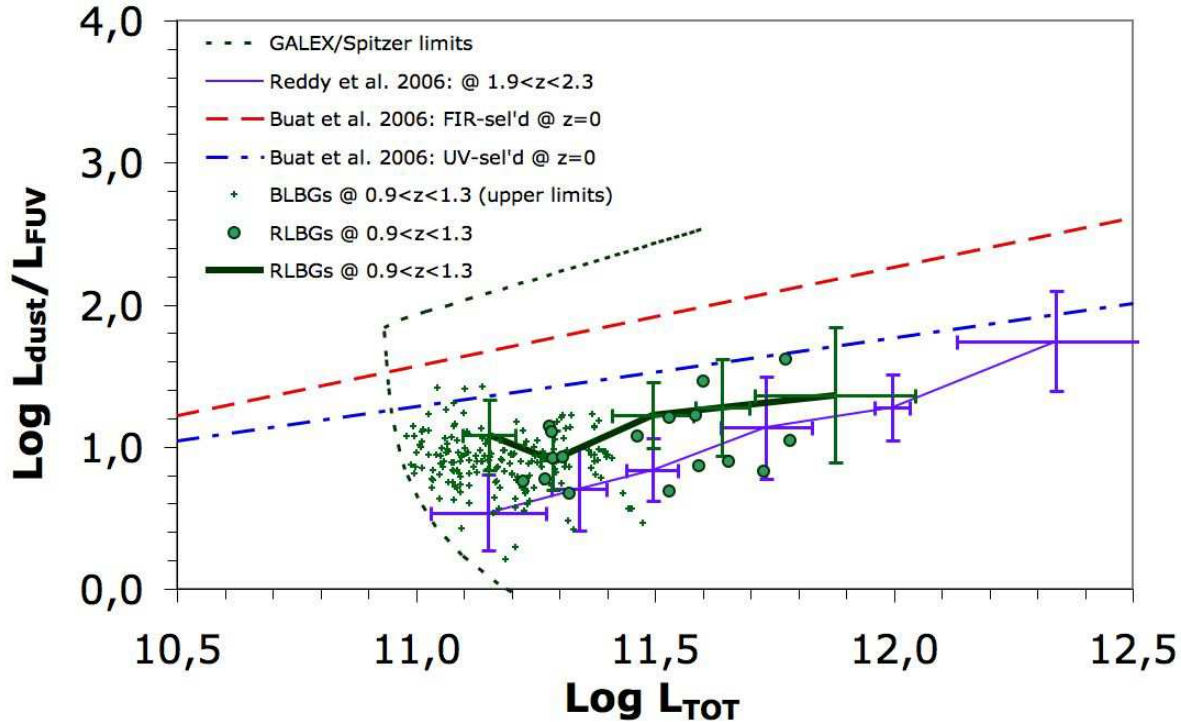}
\caption{Evolution of $L_{dust}/L_{FUV}$ ratio for UV-selected samples as a function of the redshift from 
$z \sim 0$ (Buat et al. 2005) to $z \sim 1$ (this sample: filled dots, heavy continuous line and small crosses) 
and to $z \sim 2$ (Reddy et al. 2006: thin solid line). At $z = 0$, the linear regression follows the same range 
in $Log (L_{TOT})$ covered by the FIR selection (dashed line) in Buat et al., whereas the UV selection 
(dot-dashed line) is the linear regression extrapolated beyond the observed limit at $Log (L_{TOT}) \approx 11.2$. 
The dotted line indicates the boundaries above which we cannot observe LBGs because of the GALEX and {\it Spitzer} 
sensitivity limits. Even though these boundaries limit the $Log (L_{TOT})$ range (small crosses), they should not 
have any effect on the $L_{dust}/L_{FUV}$ range.\label{fig6}}
\end{figure*}

\begin{figure}
%\vspace{2pt}
\epsfxsize=8truecm\epsfbox{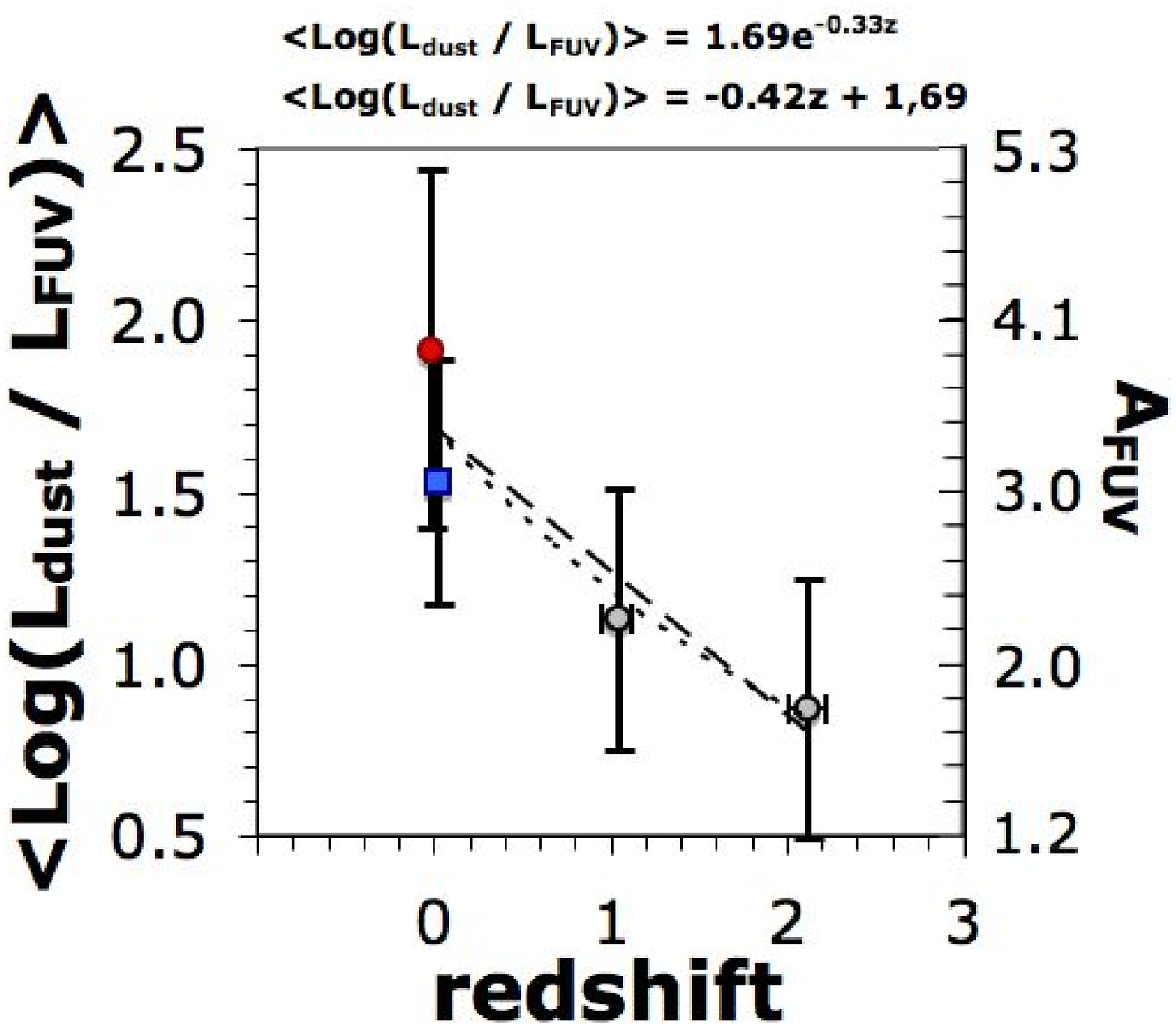}
\caption{From the present data, we seem to observe a regular decrease of $L_{TIR}/L_{FUV}$ with redshift 
from $z = 0$ (Buat et al. (2007a) to $z = 2$ (Reddy et al. (2006). Both an exponentially decaying law 
(dotted line) and a linear regression (dashed line) are a fair representation of the decline of the FUV 
dust attenuation in this redshift range.\label{fig7}}
\end{figure}

\section{The Detectability of Lyman Break Galaxies}

Will we be able to detect more UV-selected galaxies and LBGs at high redshift in the future? The main 
limitation for deep observations in the FIR are due to (i) thermal noise, (ii) cirrus and (iii) confusion 
limits (Helou \& Beichman 1990, Dole et al. 2004, Kiss, Klaas \& Lemke 2005). Using the information 
available on the next generation of FIR/submm instruments, we can assess whether they would be able to 
detect LBGs. In Fig. 8, we plot the average SED of RLBGs and BLBGs as determined in this paper (Table 4). 

On top of our two template SEDs, the flux density limits of several telescopes are also drawn to evaluate 
the detectability of a larger sample of $ z \sim 1 - 3$ LBGs/UV-selected galaxies in the future. The 
comparison is based on a 1h exposure time with a $5 \sigma$ detection for all the instruments. The Atacama 
Large Millimeter Array (ALMA) can detect RLBGs up to $z = 2.5$ in 1h and can just detect BLBGs at $z = 1$. 
A positive K-correction applies in the millimeter range but those objects are fainter even at low redshift 
and the K-correction does not help. A longer exposure time is necessary for detections at z $>$ 1. 
The Herschel space telescope 
can detect the brightest RLBGs around $100 \mu m$ but will not be able to detect fainter ones or higher 
redshift ones in 1h. Deeper exposure times are also needed to broaden the spectral range. However, 
important information will be provided at the tip of the dust emission SED to further constrain the dust 
temperatures and TIR luminosities. The European SPICA Instrument 
(ESI) on the Japanese-led cryogenically-cooled telescope SPICA can take advantage of the low background to 
benefit from the low confusion-limit for a 4-m FIR telescope. We assume, here, two kinds of detectors 
depending on which one will be available for ESI in the mid-2010 frame: Spitzer- or Herschel-like photo-detectors (GOAL) 
or state-of-the-art photodetectors (SOAP) utilizing bolometers with transition edge superconductor temperature
sensors (TES) (Swinyard et al. 2006). ESI would do a fair  job in 1h below $80 \mu m$ for photoconductors and 
reach exceptional sensitivies 
for the TES-based devices. In the first case, it would provide us with individual data on $z = 1$ BLBGs 
while in the second 
case, all LBGs would be detected up to $z = 2.0$ in 1h. Finally on the blue side of the dust emission 
JWST/MIRI would be able to detect the mid-IR emission of low-redshift LBGs (z of 1 - 2) but higher redshift 
ones would be below the detection limits as the aromatic features shift out of the longest photometric band.

\begin{figure*}
%\vspace{2pt}
\epsfxsize=16truecm\epsfbox{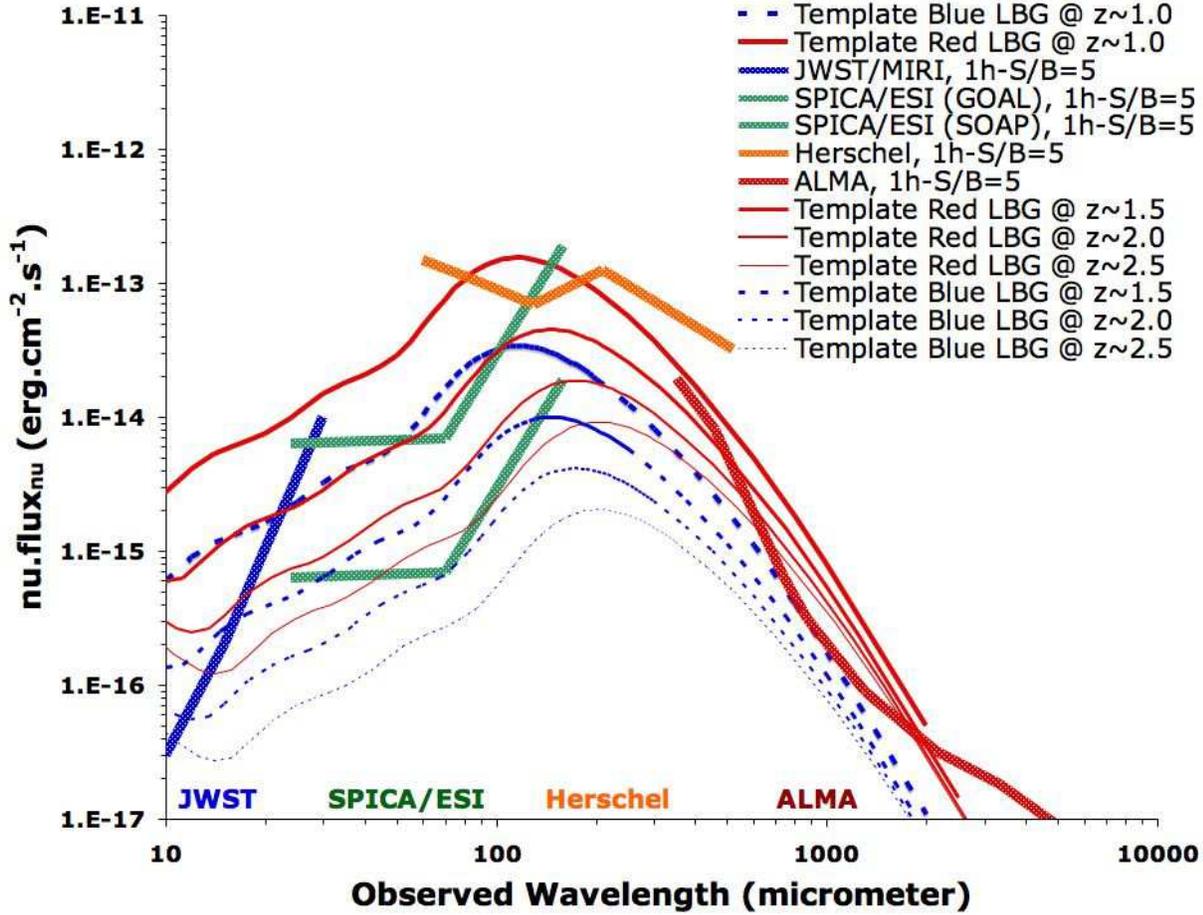}
\caption{The expected performances of JWST, SPICA/ESI, Herschel and ALMA are overplotted (thick lines with ranges 
of corresponding facilities quoted at the bottom of the figure) in this diagram over our template SED for BLBGs 
(solid lines, the higher the redshift the thinner the line) and for RLBG (dashed lines, the thinner the redshift 
the thinner the line). GOAL uses the characteristics of presently available photoconductors (e.g. Herschel) 
whereas the State Of the Art Photodetectors (SOAP) may be available  in the mid-2010 time frame.\label{fig8}}
\end{figure*}

\begin{table}
\centering
%\begin{minipage}{90mm}
\caption{Template SEDs (in units of erg.cm$^{-2}$.s$^{-1}$) of RLBGs and BLBGs at $z \sim 1$. We assume that 
1) all RLBGs share the same flux density at 70 $\mu m$, 2) that the Rayleigh-Jeans part of the SED can be well 
represented by the SED of HR10 from Stern et al. (2006) both for RLBGs and BLBGs and 3) we use the flux density 
evaluated in Burgarella et al. (2006b) by stacking the 24 $\mu m$ images for  BLBGs. }
\begin{tabular}{@{}ccccc@{}}
\hline
$\lambda_{rest}~(\mu m)$ & $\nu.f_\nu$(RLBG) & $\nu.f_\nu$(BLBG) \\
%                        & (erg.cm$^{-2}$.s$^{-1}$)  & (erg.cm$^{-2}$.s$^{-1}$) & (erg.cm$^{-2}$.s$^{-1}$)  & (erg.cm$^{-2}$.s$^{-1}$) \\
\\
0.1 & 8.09E-15 & 6.15E-15 \\
0.2 & 1.36E-14 & 6.94E-15 \\
0.3 & 1.26E-14 & 5.35E-15 \\
0.4 & 2.10E-14 & 7.30E-15 \\
0.5 & 1.88E-14 & 5.98E-15 \\
0.6 & 1.72E-14 & 5.13E-15 \\
0.7 & 1.59E-14 & 4.50E-15 \\
0.8 & 1.50E-14 & 4.11E-15 \\
0.9 & 1.45E-14 & 3.86E-15 \\
1.0 & 1.42E-14 & 3.70E-15 \\
2.0 & 7.13E-15 & 1.69E-15 \\
3.0 & 2.86E-15 & 6.61E-16 \\
4.0 & 2.05E-15 & 4.56E-16 \\
5.0 & 2.72E-15 & 5.91E-16 \\
6.0 & 4.10E-15 & 8.87E-16 \\
7.0 & 5.25E-15 & 1.13E-15 \\
8.0 & 6.04E-15 & 1.30E-16 \\
9.0 & 6.70E-15 & 1.45E-16 \\
10.0 & 7.49E-15 & 1.62E-15 \\
12.0 & 9.87E-15 & 2.13E-15 \\
12.5 & 1.06E-14 & 2.30E-15 \\
15.0 & 1.48E-14 & 3.19E-15 \\
17.5 & 1.82E-14 & 3.93E-15 \\
20.0 & 2.08E-14 & 4.49E-15 \\
22.5 & 2.39E-14 & 5.15E-15 \\
25.0 & 2.88E-14 & 6.22E-15 \\
27.5 & 3.67E-14 & 7.92E-15 \\
30.0 & 4.76E-14 & 1.03E-14 \\
32.5 & 6.09E-14 & 1.31E-14 \\
35.0 & 7.56E-14 & 1.63E-14 \\
37.5 & 9.07E-14 & 1.96E-14 \\
40.0 & 1.05E-13 & 2.27E-14 \\
42.5 & 1.18E-13 & 2.55E-14 \\

\hline
\end{tabular}
%\end{minipage}
\end{table}

\begin{table}
\centering
%\begin{minipage}{90mm}
\caption{Table 7 continued}
\begin{tabular}{@{}ccccc@{}}
\hline
$\lambda_{rest}~(\mu m)$ & $\nu.f_\nu$(RLBG) & $\nu.f_\nu$(BLBG) \\
%                        & (erg.cm$^{-2}$.s$^{-1}$)  & (erg.cm$^{-2}$.s$^{-1}$) & (erg.cm$^{-2}$.s$^{-1}$)  & (erg.cm$^{-2}$.s$^{-1}$) \\
\\
45.0 & 1.29-13 & 2.79E-14 \\
47.5 & 1.38E-13 & 2.99E-14 \\
50.0 & 1.45E-13 & 3.14E-14 \\
52.5 & 1.50E-13 & 3.24E-14 \\
55.0 & 1.53E-13 & 3.30E-14 \\
57.5 & 1.54E-13 & 3.33E-14 \\
60.0 & 1.54E-13 & 3.32E-14 \\
62.5 & 1.52E-13 & 3.29E-14 \\
65.0 & 1.50E-13 & 3.23E-14 \\
67.5 & 1.47E-13 & 3.16E-14 \\
70.0 & 1.43E-13 & 3.08E-14 \\
72.5 & 1.38E-13 & 2.99E-14 \\
75.0 & 1.34E-13 & 2.89E-14 \\
77.5 & 1.29E-13 & 2.79E-14 \\
80.0 & 1.24E-13 & 2.68E-14 \\
82.5 & 1.19E-13 & 2.58E-14 \\
85.0 & 1.15E-13 & 2.47E-14 \\
87.5 & 1.10E-13 & 2.37E-14 \\
90.0 & 1.05E-13 & 2.27E-14 \\
92.5 & 1.01E-13 & 2.17E-14 \\
95.0 & 9.62E-14 & 2.08E-14 \\
97.5 & 9.20E-14 & 1.99E-14 \\
100.0 & 8.80E-14 & 1.90E-14 \\
125.0 & 5.64E-14 & 1.22E-14 \\
150.0 & 3.71E-14 & 8.00E-15 \\
200.0 & 1.75E-14 & 3.77E-15 \\
300.0 & 5.01E-15 & 1.08E-15 \\
400.0 & 1.84E-15 & 3.96E-16 \\
500.0 & 8.00E-16 & 1.73E-16 \\
1000.0 & 4.95E-17 & 1.07-17 \\
\hline
\end{tabular}
%\end{minipage}
\end{table}

\section{Summary}

We applied a PSF-fitting method to estimate the flux of a sample of NUV=26.2 UV-selected objects in the GALEX Deep 
Imaging Survey of the Chandra Deep Field South. The GALEX deep images start to be confusion-limited ($\sim$ 20 beams 
per source) and fluxes of objects fainter than $NUV \approx 22$ are over-estimated by the GALEX pipeline because 
objects can be blended. The selection of galaxies through a FUV dropout provides a very complete sample 
of star-forming galaxies in the redshift range $0.9 \le z \le 1.3$. However, the efficiency is low as 
compared to a color-color method (e.g. Adelberger et al. 2004).

We analysed the spectral energy distributions and the luminosity function of this sample of $z \sim 1$ LBGs,
using both $GALEX$ and $Spitzer/MIPS$ data. We divide the sample into two sub-classes depending 
on whether they are detected 
at 24$\mu m$ (red LBGs, RLBGs) or are not (blue LBGs, BLBGs). 
The NIR/MIR part of both SEDs indicates that AGNs are unlikely to play a major role in these objects; they
are dominated by star formation. The UV to NIR part of the two template SEDs can be fitted by PEGASE 
models with exponentially 
decaying star formation histories of ages 500 Myrs and 250 Myrs for RLBGs and BLBGs respectively and the same dust 
attenuation law proportional to $(\lambda / \lambda_{FUV)})^{-0.7}$. The only difference comes from 
the amount of dust attenuation: $A_{FUV} = 2.5$, constrained by the $L_{TIR}/L_{FUV}$ for RLBGs and 
$A_{FUV} = 1.8$ for BLBGs. 
The mean $L_{TIR} = 10^{11.5}$ for RLBGs (i.e. LIRGs) and  $L_{TIR} = 10^{10.9}$ for BLBGs. 

The BLBG SED template at $z \sim 1$ has a shape similar to the mean SED of LBGs at $z \sim 3$ published by 
F\"orster-Schreiber et al. (2004). This means that the latter objects undergo low dust attenuation, as expected 
from the selection. However, observations by Huang et al. (2005) suggest that a population of dusty LBGs 
with large total SFRs also exists at high redshift. Such a population appears to be less numerous than the BLBGs, 
as is also the case at $z \sim 1$. On the one hand, we find that the summed UV star formation rate 
(uncorrected for dust attenuation) of our LBGs 
only represents $\sim 1/3$ of the total (i.e. UV + dust) star formation rate of all LBGs. That implies a large 
and uncertain extrapolation to get the total star formation density.  On the other hand BLBGs, not detected 
in FIR, form the bulk (in number) of the galaxy population selected in UV.

Our LBG sample at $z \sim 1$ has a UV luminosity function consistent with that of purely UV-selected galaxies. 
The similarity of the two luminosity functions implies that making use of color selection is 
not useful to study star-forming galaxies whenever redshifts are available.
UV measurements corrected using th $IRX-\beta$ method provide correct (on average) but relatively uncertain 
total SFRs, even for our LBG / LIRG sample. The dispersion is larger than an estimate involving both UV 
and IR measurements. For RLBGs, $SFR_{TIR}$ provides the a reasonably accurate estimate of $SFR_{TOT}$.

Finally, we observe an apparent regular decrease of the ratio $L_{TIR} / L_{FUV}$ for UV-selected galaxies 
from $z = 0$ to $z \approx 2$, using Buat et al. (2005) and Reddy et al. (2006) values in addition to ours. 
If confirmed, this trend might have strong cosmological implications in terms of the star formation history 
of the universe and the timescales for dust formation in primordial galaxies.

\section*{Acknowledgments}

DB and VB thank the French Programme National Galaxies and Programme National de Cosmologie for financial 
support. TTT has been supported by the 21st Century COE Program "Exploring New Science by Bridging 
Particle-Matter Hierarchy", Tohoku University. TTT has also been supported by the Special Coordination 
Funds for promotion Science and Technology (SCF). This work was also partially supported by contract
1255094 from JPL/Caltech to the University of Arizona.

\bsp

\label{lastpage}

\end{document}